\documentclass[a4paper,11pt]{article}
\usepackage{pos}

\title{Lepton Flavor in Composite Higgs Models}

\author*[a]{Florian Goertz}

\affiliation[a]{Max-Planck-Institut f{\"u}r Kernphysik,\\ Saupfercheckweg 1, 69117 Heidelberg, Germany}

\emailAdd{fgoertz@mpi-hd.mpg.de}

\abstract{In this talk, I will present the status of lepton flavor physics in composite Higgs models with partial compositeness in the light of recent data in the lepton sector. I will consider anarchic flavor setups, scenarios with flavor symmetries, and minimal incarnations of the see-saw mechanism that naturally predict non-negligible lepton compositeness.
The focus will be on lepton flavor violating processes, dipole moments, and on probes of lepton flavor universality, all providing stringent tests of partial compositeness. The expected size of effects in the different approaches to lepton flavor and the corresponding constraints will be discussed, including ‘UV complete’, effective, and holographic descriptions.}

\FullConference{%
  *** Particles and Nuclei International Conference - PANIC2021 ***\\
  *** 5 - 10 September, 2021 ***\\
  *** Online *** 
}

\begin{document}
\setlength{\textheight}{25.19cm}
\maketitle

\setlength{\topmargin}{1cm}
\section{Introduction}
\label{sec:intro}
\vspace{-2.8mm}
The concept of partial compositeness (PC)~\cite{Kaplan:1991dc,Agashe:2004rs,Contino:2003ve,Contino:2006qr} in composite Higgs (CH) models offers an attractive means to explain the flavor hierarchies. In such a framework, the Higgs is 
a composite pseudo Nambu-Goldstone boson (pNGB) of a spontaneously broken global symmetry of a new confining sector. Linear mixings of the Standard Model~(SM) fermions with composite resonances can then address the observed mass spectra. At the same time, dangerous flavor-chaning neutral currents (FCNCs) are suppressed by a 'geometric' GIM mechanism~\cite{Gherghetta:2000qt,Huber:2000ie,Agashe:2003zs,Agashe:2004cp,Agashe:2004rs}. 

In spite of obvious similarities, the lepton sector is distinct from the quark sector due to
the leptonic mixing angles being sizable \cite{Zyla:2020zbs} and all leptons being significantly lighter than the weak scale [with neutrino masses residing even many orders of magnitude below the charged spectrum, $m_{\nu_i}\!\lesssim\! 1$\,eV]. Thus, one could expect that leptons are largely elementary and should behave SM-like, showing negligible effects on the one-loop Higgs potential
or in not extremely well-measured lepton observables.
However, as we will show, different explicit models addressing the lepton masses and mixings predict some
lepton compositeness, changing the picture. Moreover, stringent flavor limits can lead to relevant bounds even for mostly elementary leptons. 

PC for leptons has been considered in \cite{Huber:2001ug,Huber:2002gp,Huber:2003tu,Moreau:2005kz,Agashe:2006iy,Perez:2008ee,Csaki:2008qq,Agashe:2008fe,Agashe:2009tu,Chen:2009gy,delAguila:2010vg,Kadosh:2010rm,Csaki:2010aj,delAguila:2010es,Hagedorn:2011un,Hagedorn:2011pw,KerenZur:2012fr,vonGersdorff:2012tt,Kadosh:2013nra,Ding:2013eca,Redi:2013pga,Frank:2014aca,Carmona:2014iwa,Chen:2015jta,Carmona:2015ena,Carmona:2016mjr,Frigerio:2018uwx}, mostly from a low-energy perspective and via holographic methods. UV completions have been envisaged, too, considering the fundamental constituents and the dynamics forming the bound states that
mix with the elementary SM-like fermions \cite{Barnard:2013zea,Ferretti:2013kya,Ferretti:2014qta,Vecchi:2015fma,Sannino:2016sfx,Cacciapaglia:2017cdi,Agugliaro:2019wtf} (see also \cite{Cacciapaglia:2019dsq,Cacciapaglia:2020jvj}).\
A subclass, dubbed "fundamental partial compositeness"~(FPC)~\cite{Sannino:2016sfx,Cacciapaglia:2017cdi,Agugliaro:2019wtf}, assumes the composite fermions to consist of an elementary fermion ${\cal F}$ and a scalar ${\cal S}$. In contrast to 3-fermion states, the scaling dimension of the composite operator ${\cal O}_F$ is then expected close to the canonical $[{\cal O}_F]_0\! =\![{\cal F}]\!+\![{\cal S}]\!=\!5/2\! \ll\! 3 [{\cal F}]\!=\!9/2$, allowing for viable masses via modest anomalous dimensions, see below (the scalars could also emerge ultimately from fermions~at~higher~scales). 
\vspace*{-2mm}

\section{Partial Compositeness}
\vspace{-3mm}
\paragraph{UV Realization of Partially Composite Leptons}
Assuming the composite operators 
${\cal O}_F\!=\!\langle {\cal F S} \rangle$, explicit models of FPC have been presented, employing the confining technicolor (TC)-like groups $G_{\rm TC} =$ SU($N_{\rm TC}$), SO($N_{\rm TC}$), Sp($N_{\rm TC}$) \cite{Sannino:2016sfx,Sannino:2017utc,Cacciapaglia:2017cdi}. Below, we focus on $G_{\rm TC}=$ Sp($N_{\rm TC}$) with four Weyl fermions ${\cal F}^a, a=1,..,4$, per TC, resulting in the global symmetry breaking pattern SU(4)$_{\cal F}$ $\to$ Sp(4)$_{\cal F}$ after formation of the condensate~\cite{Cacciapaglia:2017cdi,Sannino:2017utc} (with $\epsilon_{\rm TC}$ the antisymmetric~$G_{\rm TC}$~tensor)
\vspace*{-0.3mm}
\begin{equation}
\langle {\cal F}^a \epsilon_{\rm TC}\, {\cal F}^b \rangle
= \Lambda_c f^2 \Sigma_\theta^{ab} \,.
\end{equation}
Here $a,\!b\!\in\,$SU(4)$_{\cal F}$, the global symmetry of the (massless, $m_{\cal F}\!=\!0$) techni-fermions, which contains the SM electroweak (EW) group, with $F_\alpha\!=\!({\cal F}_1,{\cal F}_2)^T$ a doublet without hypercharge $Y\!=\!0$ and ${\cal F}_{3,4}$ singlets with $Y\!=\!\mp 1/2$, respectively~\cite{Cacciapaglia:2017cdi,Sannino:2017utc}. Moreover, $\Lambda_c\!\sim\!4\pi f$ is the TC condensation scale (with $f$ the pNGB decay constant), and $\Sigma_\theta = \cos\theta\ {\rm diag}(i \sigma_2,\!-i \sigma_2) + \sin\theta\ {\rm offdiag}(\bf{1}_2,\!-\bf{1}_2)$ parametrizes the vacuum-alignment with the Sp(4)$_{\cal F}$ group, with the EW vacuum expectation value (vev) $v\!=\!\sin \theta f\!$, such that $\sin\theta\!=\!0$  $(\sin\theta\!=\!1)$ corresponds to unbroken (fully broken) EW symmetry.
Finally, the pNGBs of the SU(4)$_{\cal F}/$Sp(4)$_{\cal F}$ coset (with generators $T_\theta^{\hat a}$) are parameterized as fluctuations around this vacuum via the Goldstone matrix $
\Sigma\!= \exp(i2\!\sqrt 2/\!f \, \Pi_{\hat a} T_\theta^{\hat a}) \Sigma_\theta,$
where $\Pi_{1,2,3}$ are the EW Goldstones, $\Pi_4$ is the Higgs, and $\Pi_5$ a new singlet.

Moreover, 12 complex scalars are added to realize FPC with $[{\cal O}_F]\! \approx\! 5/2$,
forming {\it 3 generations} of color triplets ${\cal S}_q$
with $Y\!=\!-1/6$, and corresponding color singlets ${\cal S}_l$ with $Y\!=\!1/2$. Defining ${\cal S}=({\cal S}_q,{\cal S}_l)^T$, the full kinetic and mass terms of the TC sector are given by
$
{\cal L}_{\rm kin}=-\frac 1 4 {\cal G}_{\mu\nu}^a {\cal G}^{a\,\mu\nu} + i {\cal F}^\dagger \bar \sigma^\mu D_\mu {\cal F} - \left({\cal F}^T \frac{m_{\cal F}}{2} \epsilon_{\rm TC} {\cal F} + {\rm h.c.} \right) +(D_\mu {\cal S})^\dagger (D^\mu {\cal S}) -{\cal S}^\dagger m_{\cal S}^2 {\cal S},\, 
$
describing the TC gauge, fermion, and scalar sectors, with the last one exhibiting a global Sp(24)$_{\cal S}$ symmetry for $m_{\cal S}\!=\!0$, with $\Phi\!=\!({\cal S},-\epsilon_{TC} {\cal S}^\ast)^T$ in the fundamental representation. 
This setup is called "minimal FPC" (MFPC) -- featuring a minimal amount of techni-matter that leads to a pNGB Higgs and fermionic resonances mixed linearly with the~SM~\cite{Sannino:2016sfx,Cacciapaglia:2017cdi,Sannino:2017utc}.
Embedding the SM-like fermions into  $24_{\cal S} \otimes \bar 4_{\cal F}$ of the full global symmetry (the former containing color and flavor) we can couple them to the TC sector, while respecting the SM gauge group,
and the resulting Lagrangian for leptons reads~\cite{Cacciapaglia:2017cdi,Sannino:2017utc}
\vspace*{-1.5mm}
\begin{equation}
\label{eq:PCF}
{\cal L}_{\rm yuk} \supset  L_\alpha\, y_L\, \Big({\cal S}_l \epsilon_{\rm TC} F^\alpha\Big) + \bar e \, y_{\bar e} \, \left({\cal S}^\ast_l  \bar {\cal F}^3\right) -  \bar \nu \left( \tilde y_{\bar \nu} \left({\cal S}_l  \bar{\cal F}^3\right) +y_{\bar \nu}\left({\cal S}^\ast_l  \bar {\cal F}^4\right)\right) + {\rm h.c.}\,.
\end{equation} This provides a UV realization of PC, that can leave an imprint in low energy observables,~see~below.

\paragraph{Effective Description and Different Embeddings of the Lepton Sector}
To capture a broader range of potential UV completions for the phenomenological study and to make contact with a large set of previous works on lepton flavor -- which were performed in an effective/holographic approach, often employing the SO(5)/SO(4) coset -- in the following we will consider an effective description of PC.
This allows us to describe broad features of PC realizations via different structures of mixing coefficients at low energies. Below the scale $\Lambda_{\rm UV}$ where the elementary/composite-sector interactions are generated (Eq.~(\ref{eq:PCF}) for MFPC), PC for leptons can be described via linear mixings of the SM-like elementary fields with composite operators ${\cal O}^l_{L,R}$ of the confining sector
\vspace*{-1mm}
\begin{equation}
\label{eq:PC}
{\cal L}_{\rm mix} = (\lambda_L^\ell/\Lambda_{\rm UV}^{\gamma^\ell_L})\, \bar l^\ell_L {\cal O}^\ell_L + (\lambda_R^\ell/\Lambda_{\rm UV}^{\gamma^\ell_R})\, \bar l^\ell_R {\cal O}^\ell_R + (\lambda_R^{\nu_\ell}/\Lambda_{\rm UV}^{\gamma^{\nu_\ell}_R})\, \bar \nu^\ell_R {\cal O}^{\nu_\ell}_R \, \, +{\rm h.c.},
\end{equation}
eventually responsible for the lepton masses. Here, $\ell=e,\mu,\tau$, with an obvious analogue for quarks, $\lambda_{L,R}^l$ are (${\cal O}(1)$) couplings and $\gamma^l_{L,R} = [{\cal O}^l_{L,R}]-5/2$ are the anomalous dimensions of~the~composite-sector operators. Moreover, $l^\ell_L$, $l^\ell_R$, and $\nu_R^\ell$ correspond to the embeddings of the SM-like fields into irreducible representations of the global symmetry of the composite sector, such as~SO(5)~in the Minimal Composite Higgs Model (MCHM) \cite{Agashe:2004rs}, according to the operators they mix with. 


Small differences in $\gamma^i_{R,L}$ lead to hierarchical fermion masses (and possibly mixings)  \cite{Casagrande:2008hr,Csaki:2008zd,Huber:2000ie,Huber:2003tu} from an anarchic UV structure, after integrating out the resonances excited by ${\cal O}^l_{L,R}$, inducing
\vspace*{-2.3mm}
\begin{equation}
\label{eq:ml}
m_l \sim g_\ast v/\sqrt{2}\, \epsilon_L^l \epsilon_R^l\,, \quad\text{where }\, \epsilon^l_{L,R}\!\sim \lambda_{L,R}^{l}/g_\ast\ (\mu/\Lambda_{\rm UV})^{\gamma_{L,R}^{l}}
\vspace*{-1mm}
\end{equation}
defines the 'degree of compositeness' of a chiral SM-like field. Here, $g_\ast$ is the coupling~of~the~resonances and $\mu\!\sim\!\Lambda_c\!\sim\!\rm{TeV}$ the IR scale where the composites condense (see \cite{CHReview} for~more~details).\footnote{It would be interesting to examine the emergence of a hierarchical spectrum explicitly in a setup of MFPC.}


\vspace*{-1.2mm}
\subparagraph{Basic Anarchic Setup}
Similar to the (frequently studied) quark sector, leptons in CH models can be realized just by assuming anarchic values for the dimensionless input parameters, generating the hierarchies in charged-lepton masses after condensation by the UV-scale suppression, $\Lambda_{UV}\!\gg\!\mu$ in Eq.~(\ref{eq:PC}). The leptonic mixing matrix and neutrino masses could be kept non-hierarchical by appropriate assumptions on the PC structure, as envisaged in Refs~\cite{Agashe:2006iy,Agashe:2008fe,Agashe:2009tu,Csaki:2010aj,KerenZur:2012fr,Frigerio:2018uwx}. However, even though there is a suppression of FCNCs from PC, it remains a challenge to evade the stringent flavor constraints, as we will see, pushing $f$ above the TeV scale.

\vspace*{-1.2mm}
\subparagraph{Models with Flavor Symmetries}
Flavor symmetries can be used to refine the anarchic approach, properly generating the particular form of the leptonic mixing matrix and of neutrino masses together with hierarchical charged-lepton masses and a flavor protection going beyond the geometric GIM, see below.
Popular flavor groups~$G_f$ are summarized in Tab.~\ref{tab:Gf}. The discovery of a non-zero $\theta_{13}$ angle
\cite{Abe:2011sj,An:2012eh,Ahn:2012nd,Adamson:2013whj} lead to a broadening to (product) groups beyond the early $A_4,S_4$, or (double tetrahedral) $T^\prime$, and to considering spontaneous breaking of such symmetries.
Interestingly, models with flavor symmetries often feature a suppression of the Yukawa couplings in the composite sector (inducing lepton masses after mixing), since those control the breaking of $G_f$ (see, e.g., \cite{delAguila:2010vg}). Moreover, left-handed (LH) lepton compositeness is bound to be small due to lack of custodial protection of $Z \bar \ell_L \ell_L$ couplings. In turn, the $\tau_R$ needs to mix quite significantly with the composite sector to generate $m_\tau$, which can lead to interesting LHC/Higgs signatures~\cite{delAguila:2010es,Carmona:2013cq,Carmona:2013lva,Carmona:2014iwa}.
Moreover, non-negligible compositeness in the charged lepton sector can also emerge beyond such models of flavor symmetries, simply from the scale of neutrino masses -- as in minimal realizations of a seesaw in the CH framework that we will discuss now.

\begin{table}[t!]
\caption{Popular choices for flavor symmetry $G_f$ in the lepton sector, where below $X \in \{A_5, \Delta(96),\Delta(384)\}$.\vspace{-1.8mm}}
\centering
\label{tab:Gf}      
\begin{tabular}{l||l|l|l|l|l|l|l}
$G_f$ & $A_4 \times Z_N$  & $S_4 \!\times\! Z_N^n$ &  $X \!\times\!  Z_N$ & $\Delta(27) \!\times\! Z_4\!\times\! Z_4^\prime$ & $S_3$ & $T^\prime$ & $U(N)$ \\
\noalign{\smallskip}\hline\noalign{\smallskip}
Ref.   & \cite{Csaki:2008qq,delAguila:2010vg,Kadosh:2010rm,Kadosh:2013nra} & \cite{Hagedorn:2011un,Hagedorn:2011pw,Ding:2013eca} & \quad
\cite{Hagedorn:2011pw}   & \quad 
\cite{Chen:2015jta} & \cite{Frank:2014aca} &\cite{Chen:2009gy}  &  \cite{vonGersdorff:2012tt,Frigerio:2018uwx}\\
\end{tabular}
\vspace{-2.6mm}
\end{table}

\vspace*{-1.5mm}
\subparagraph{Minimal Seesaw Model and Composite Leptons}
In the CH framework, a very minimal realization of the lepton sector is possible that explains the tiny neutrino masses via a \mbox{type-III} seesaw with heavy fermionic SU(2)$_L$ triplets, providing at the same time an efficient flavor protection. Employing such triplets, a {\it unification} of the right-handed (RH) lepton sector is possible and a single, symmetric, representation of SO(5) can host both the charged RH leptonic SU(2)$_L$ singlet as well as the RH seesaw triplet \cite{Carmona:2015ena,Carmona:2016mjr,Carmona:2014iwa}. 
This leads to a more minimal model for leptons, featuring less new particles and less parameters, than conventional analogues to minimal viable (MCHM$_5$-like) quark sectors, which require in fact mixings with {\it four} fundamental {\bf 5}'s of SO(5)~\cite{delAguila:2010vg,Contino:2006qr}. 

Here, the PC Lagrangian~(\ref{eq:PC}) only contains linear interactions with {\it two} operators, embedding $l^\ell_L \!\sim\! {\bf 5} =\!({\bf 2},{\bf 2}) \oplus ({\bf 1},{\bf 1})$ and $l^\ell_R\! \sim\! {\bf 14}=\!({\bf 3},{\bf 3}) \oplus ({\bf 2},{\bf 2}) \oplus ({\bf 1},{\bf 1})$, where the former hosts the
LH SM doublet in the $({\bf 2},{\bf 2})$ of ${\rm SU(2)}_L\!\times\! {\rm SU(2)}_R$ and the latter contains now both the see-saw triplet~$\Sigma_{\ell}$~and RH singlet in the $({\bf 3},{\bf 3})$ and $({\bf 1},{\bf 1})$, see \cite{Carmona:2015ena,Carmona:2016mjr}. Adding a Majorana mass term in the elementary~sector
\vspace*{-2mm}
\begin{equation}
\label{eq:Maj}
		\mathcal{L}_{\rm el} = -\frac{1}{2}\left[M_{\Sigma}^{\ell \ell^{\prime}}\mathrm{Tr}\left(\bar{\Sigma}_{\ell R}^{c}\Sigma_{\ell^{\prime}R} \right)+\mathrm{h.c.} \right]\,,\quad \Sigma_{\ell}=\begin{pmatrix}\hat{\nu}^\ell/\sqrt{2}&\hat{\lambda}^\ell\\ \hat \ell& -\hat{\nu}^\ell/\sqrt{2}\end{pmatrix},\ \ell=e,\mu,\tau\,,
\vspace*{-2.3mm}
\end{equation}
explains the tiny neutrino masses via large $M_\Sigma^{\ell\ell\prime} \gg v$. If now $l_R^\ell \supset \Sigma_\ell$ in Eq.~(\ref{eq:PC}) would be fully elementary, an effective Majorana mass of ${\cal O}(M_{\rm Pl})$ - the fundamental UV scale of the theory - would emerge, resulting in a significantly too strong suppression of neutrino masses (note that also the Dirac mass is PC-suppressed).
This calls for a non-negligible compositeness of $l^\ell_R$, bringing down the effective Majorana mass~\cite{Carmona:2015ena,Carmona:2016mjr,CHReview}. A hierarchy $0 \!\ll\! \epsilon^\tau_{R}\!\ll\! \epsilon^\mu_{R}\!\ll\! \epsilon^e_{R}$ follows from the fact that charged lepton hierarchies require $\epsilon^e_{L}\!\ll\! \epsilon^\mu_{L}\!\ll\! \epsilon^\tau_{L}\!\ll\! 1$ (where the sizable RH compositeness leads to a cancellation of the corresponding $\epsilon^\ell_{R}$ exponential in (\ref{eq:ml}) \cite{Carmona:2015ena,Carmona:2016mjr,Hagedorn:2011un}) while the neutrino mass matrix, scaling as $\mathcal{M}_{\nu}\!\sim\! v g_\ast \epsilon^\ell_{L}(M_{\Sigma}^{\ell \ell^{\prime}}\!/\!\epsilon^{\ell}_{R}\epsilon^{\ell^{\prime}}_{R})^{-1}  v g_\ast \epsilon^{\ell^{\prime}}_{L}$, should be non-hierarchical.

\begin{figure}[!t]
	\begin{center}
			\includegraphics[width=0.46\textwidth]{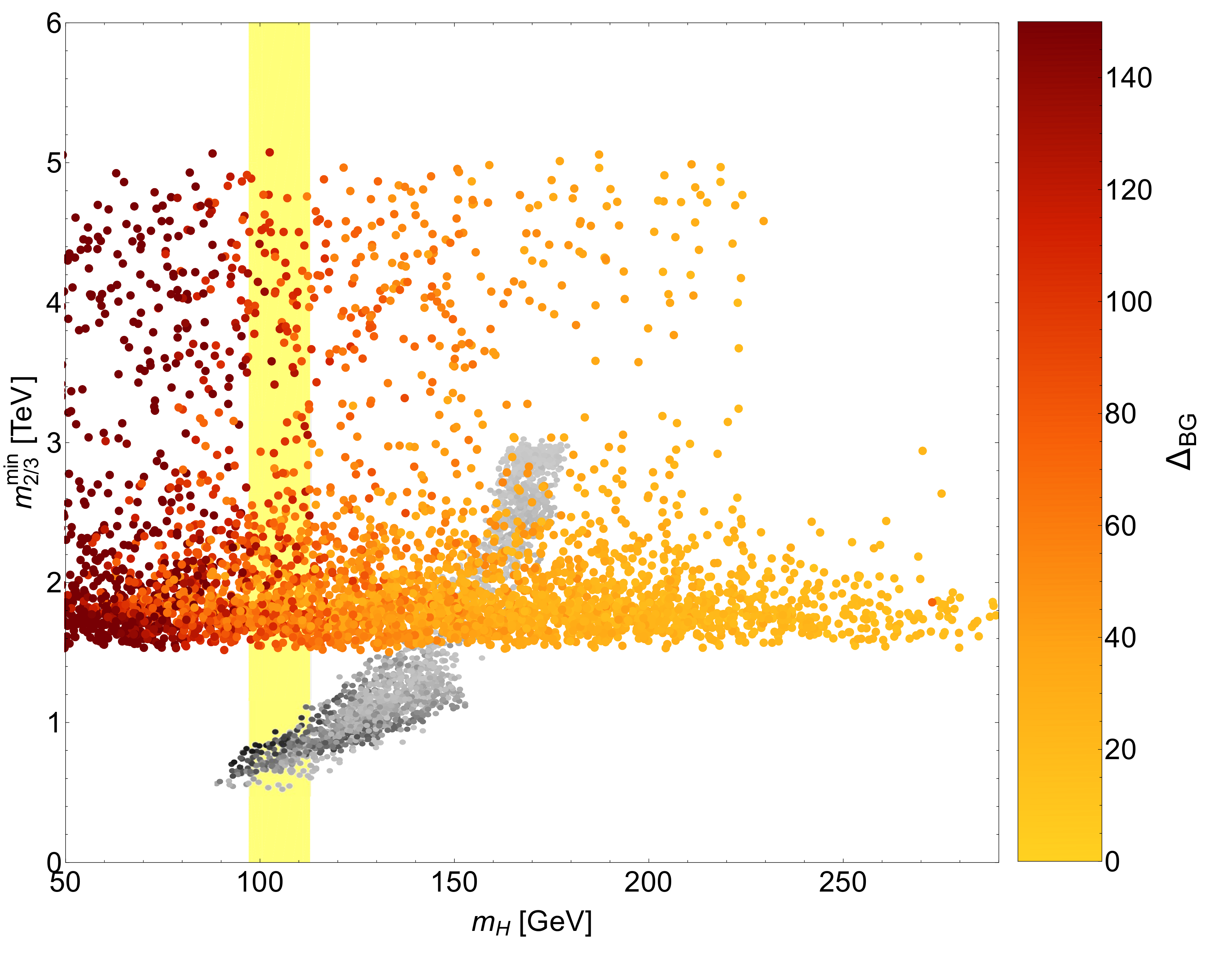}
		\ \ \	\raisebox{0.5mm}{\includegraphics[width=0.45\textwidth]{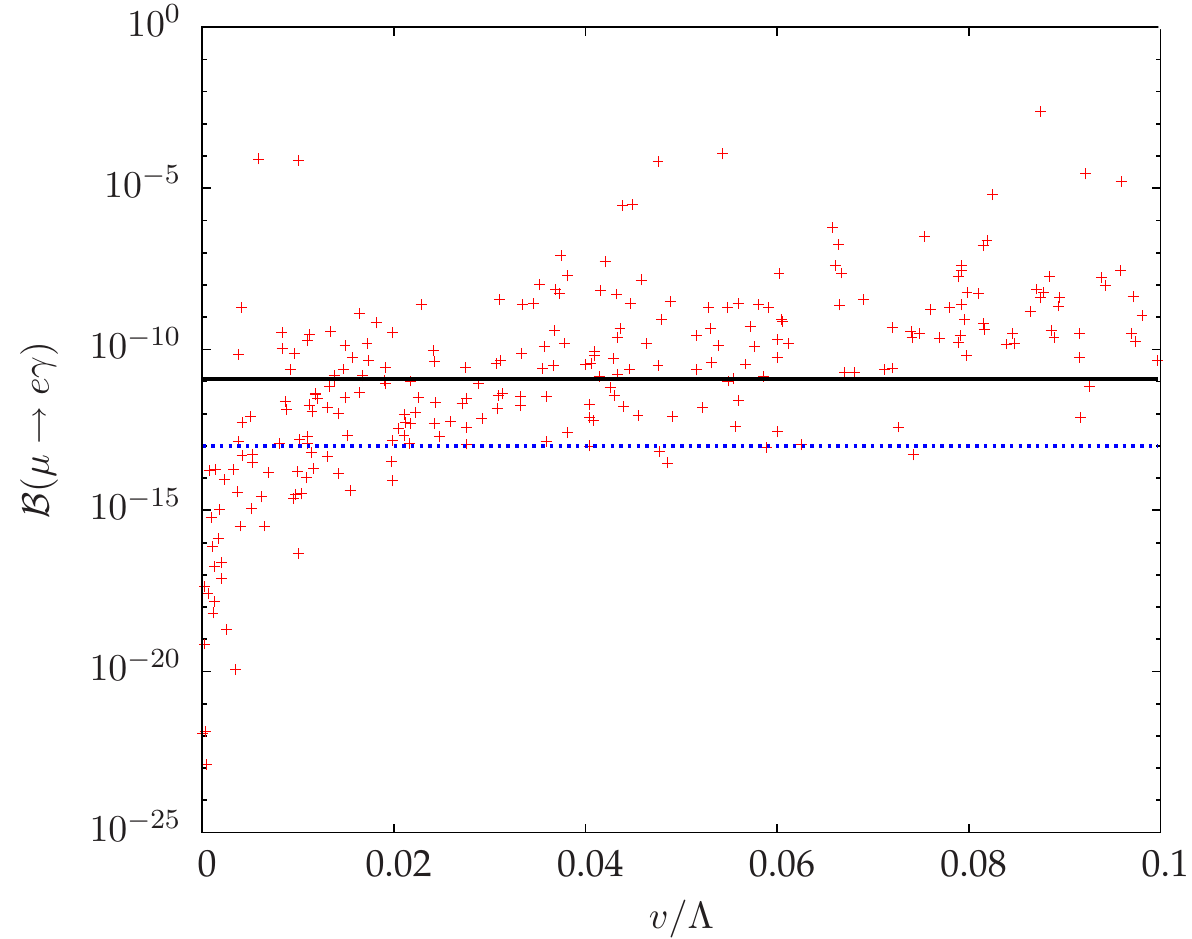}}
		 \vspace*{-3.2mm}  \caption{\footnotesize Left: Lightest top-partner mass (and tuning $\Delta_{\rm BG}$) vs.\ $m_h$ in the minimal seesaw model~(colored points) and the MCHM$_5$ (gray), for $f\!=\!1\,$TeV. Right: ${\rm BR}(\mu \to e \gamma)$ in MCHM$_5$-like setup with $A_4$ flavor symmetry vs.\
$A_4$-breaking~vev (over the cutoff) $v/\Lambda$ for $m_\ast\!\sim\! 3\,$TeV, $g_\ast\!\sim\! 4$ (from \cite{delAguila:2010vg}). The solid (dashed) line shows the MEGA (projected MEG) limit.}
			\label{fig:mhvsmtp}
		\end{center} \vspace{-6.8mm}
\end{figure}

This will lead to interesting signatures, like a non-negligible violation of lepton flavor universality (LFU) in RH couplings within the first generations,
as we will discuss below. Before, we note that the scenario is save from potentially stringent flavor and precision constraints on RH compositeness, since the minimal amount of two SO(5) representations hosting all leptons allows for a flavor symmetry broken by a {\it single} spurion in the strong sector, making it possible to diagonalize the latter~\cite{Carmona:2015ena,Carmona:2016mjr}, while custodial symmetry precisely protects the $Z \bar{\ell}_R  \ell_R$ couplings~\cite{delAguila:2010es,Carmona:2013cq}.
Moreover, the appreciable lepton compositeness, together with an (group-theoretically) enhanced contribution of the {\bf 14}, leads to a sizable leptonic impact on the pNGB Higgs mass (even for moderate compositeness)~\cite{Carmona:2014iwa}. This allows to address the significant tension of CH models, like the conventional MCHM$_5$ or MCHM$_{10}$, with bounds on top-partner masses of $m_{t^\prime} \gtrsim 1.3$\,TeV \cite{Aaboud:2018pii,Sirunyan:2018qau,Sirunyan:2019sza} by allowing for much heavier top partners while keeping $m_h$ (and $f$) low, as demonstrated in the left panel of Fig.~\ref{fig:mhvsmtp}. Here, a scan of the mass of the lightest top partner versus the Higgs mass, evaluated at $f\!=\!1\,$TeV, is performed comparing the minimal seesaw model (including a minimal quark-sector, colored points) with the MCHM$_5$ (gray points) \cite{Carmona:2014iwa,Carmona:2015ena}. While for $m_h(1\,{\rm TeV})\!\approx\! 105\,$GeV (yellow stripe) the latter is in strong tension with LHC limits, the former remains basically unconstrained (even at minimal Barbieri-Giudice tuning $\Delta_{\rm BG}$~\cite{Barbieri:1987fn}), see \cite{Carmona:2015ena,Carmona:2016mjr} and \cite{CHReview} for more details.\footnote{For a survey of (collider) constraints on CH see~\cite{Goertz:2018dyw} and for other setups addressing the top-partner issue~\cite{Panico:2012uw,Blasi:2019jqc,Blasi:2020ktl}.}

\vspace*{-2mm}
\section{Lepton Flavor: Predictions, Constraints, and Discussion}
\label{sec:PCPhen}
\vspace*{-3.25mm}

\paragraph{Lepton Flavor Violation (LFV) and Dipole Moments}
In CH models, the decay $\mu \to e \gamma$~is~induced by penguin-loop diagrams involving heavy resonances~\cite{Agashe:2006iy,Csaki:2010aj}, generating the~dipole~operator 
\vspace*{-2.95mm}
\begin{equation}
\label{eq:dip}
{\cal O}_{\ell\ell^\prime}^\gamma \equiv e v \, F_{\mu\nu} \bar\ell_{L} \sigma^{\mu\nu} \ell^\prime_{R},
\end{equation}
with here $\ell\!=\!\mu,\ell^\prime\!=\!e$.
The latest 90$\%$\,CL limit on the branching ratio (BR) reads ${\rm BR}(\mu \to e \gamma) < 4.2 \times 10^{-13}$ \cite{TheMEG:2016wtm}, being 5 orders of magnitude stronger than those on the corresponding~tau~decays~\cite{Zyla:2020zbs}. 

In the {\it anarchic scenario}, the BR scales as (see also \cite{Agashe:2013kxa,Csaki:2010aj,KerenZur:2012fr,Giudice:2007fh})
\vspace*{-1mm}
\begin{equation}
{\rm BR}(\mu \to e \gamma) = 96 \pi^2\, e^2 v^6\!/m_\mu^2 \left( \left|C^\gamma_{\mu e}\right|^2 + \left|C^\gamma_{e \mu}\right|^2 \right), \quad
C^\gamma_{\ell \ell^\prime} \sim\! \sqrt 2/32 \pi^2\ g_\ast^3/m_\ast^2\ \epsilon^\ell_L \epsilon^{\ell^\prime}_R\,,
\vspace*{-0.5mm}
\end{equation}
with $m_\ast$ the mass scale of the resonances. The experimental constraint above then leads to~the~bound
\vspace*{-1.5mm}
\begin{equation}
g_\ast^3/m_\ast^2 \sqrt{|\epsilon_L^\mu \epsilon_R^e|^2+|\epsilon_L^e \epsilon_R^\mu|^2} \lesssim 10^{-7}/{\rm TeV}^2\quad \Longrightarrow \ \,
\boxed{m_\ast/g_\ast \gtrsim 20\, {\rm TeV}\quad ({\rm BR}(\mu \to e \gamma))}\,,
\vspace*{-0.6mm}
\end{equation}
consistent with Refs~\cite{Agashe:2013kxa,Frigerio:2018uwx}, where for the second expression we employed (\ref{eq:ml}) and (conservatively) set $\epsilon_R^\ell=\epsilon_L^\ell$. This would push the anarchic scenario far beyond LHC reach, even for modest~$g_\ast\!\sim\!4$. Given new CP violation, the electric dipole moment of the electron $d_e$, being proportional to the imaginary part of the (1,1) component of the dipole coefficient of Eq.~(\ref{eq:dip}) and scaling as~\cite{Agashe:2013kxa,Frigerio:2018uwx} 
\vspace*{-2.9mm}
\begin{equation}
 d_e \sim  {\rm Im}(c_e)\, e/16\pi^2\  g_\ast^3/m_\ast^2\ \epsilon^e_{L} \epsilon^e_{R}\, v/\sqrt 2\,,
\end{equation}
where $c_e$\! contains the phase of the setup, provides another bound. 
The recent $d_e < 1.1 \times 10^{-29} e\,{\rm cm}$ from ACMEII \cite{Andreev:2018ayy} gives
\vspace*{-5.6mm}
\begin{equation}
\boxed{m_\ast/g_\ast \gtrsim \sqrt{{\rm Im}(c_e)}\, 75\, {\rm TeV}\quad ({\rm eEDM})}\,,
\end{equation}
agreeing with the limits of Refs \cite{Agashe:2013kxa,Frigerio:2018uwx}, after updating them. Finally, constraints from $\mu-e$ conversion in Gold bound FCNC couplings to the Z boson $\sim \epsilon^\mu_{L,R} \epsilon^e_{L,R}$. Setting again $\epsilon_R^\ell=\epsilon_L^\ell$, the SINDRUMII $90\%$\,CL limit of ${\rm \Gamma}(\mu Au \to e Au)/{\rm \Gamma}_{\rm capture}(\mu Au) < 7 \times 10^{-13}$ \cite{Bertl:2006up} delivers a~(weaker) bound (see \cite{Agashe:2013kxa,Frigerio:2018uwx})
\vspace*{-1.8mm}
\begin{equation}
\boxed{m_\ast/\sqrt{g_\ast} \gtrsim 3\, {\rm TeV}\quad (\mu Au \to e Au)}\,.
\end{equation}
We note that constraints from $\mu \to eee$, probing the same operator, are a factor $\sim\!3$ less~strong,~while corresponding tau decays and $d_\mu$ as well as $(g-2)_\mu$ are even less constraining, see e.g. \cite{Agashe:2013kxa,Frigerio:2018uwx}.

In setups with {\it flavor symmetries}, as in Tab.~\ref{tab:Gf}, the bounds from LFV are typically much weaker. For example, in models where leptons transform appropriately under a spontaneously broken $A_4$ symmetry~\cite{Csaki:2008qq,delAguila:2010vg}, one can rotate to a flavor-diagonal basis, up to subleading corrections, and the constraints above are reduced to the 1\,TeV scale (see also \cite{Kadosh:2013nra}). This is shown more quantitatively in the right panel of Fig.~\ref{fig:mhvsmtp}, where for small $A_4$-breaking (elementary) vev over the cutoff $v/\Lambda$ and $m_\ast\!\sim\! 3\,$TeV, $g_\ast\!\sim\! 4$, most points are in agreement with a bound of ${\rm BR}(\mu \to e \gamma)\!<\! 5\! \times\! 10^{-13}$. Similar statements hold for the {\it minimal seesaw model} with analogous flavor protection, as discussed.

In {\it MFPC},
the flavor structure in the effective theory is induced by the properties of the fundamental constituents.
If e.g. $m_{\cal S}^2\! \sim\!{\bf 1}$, as can emerge if the TC scalars acquire mass solely~from~TC interactions (and their potential conserves flavor) \cite{Sannino:2016sfx}, the coefficient of the dipole operator in Eq.~(\ref{eq:dip}) will be diagonal and real in the same basis as the SM-fermion Yukawa matrix, $C^\gamma_{\ell \ell^\prime}\!\propto\!y^{\rm SM}_{\ell \ell^\prime}$~(to leading approx.).
This will push ${\rm BR}(\mu \to e \gamma)$ and $d_e$ below the limits, even for very low TC~scale~\cite{Sannino:2016sfx}.

\begin{figure}[!t]
	\begin{center}
	\includegraphics[width=0.359\textwidth]{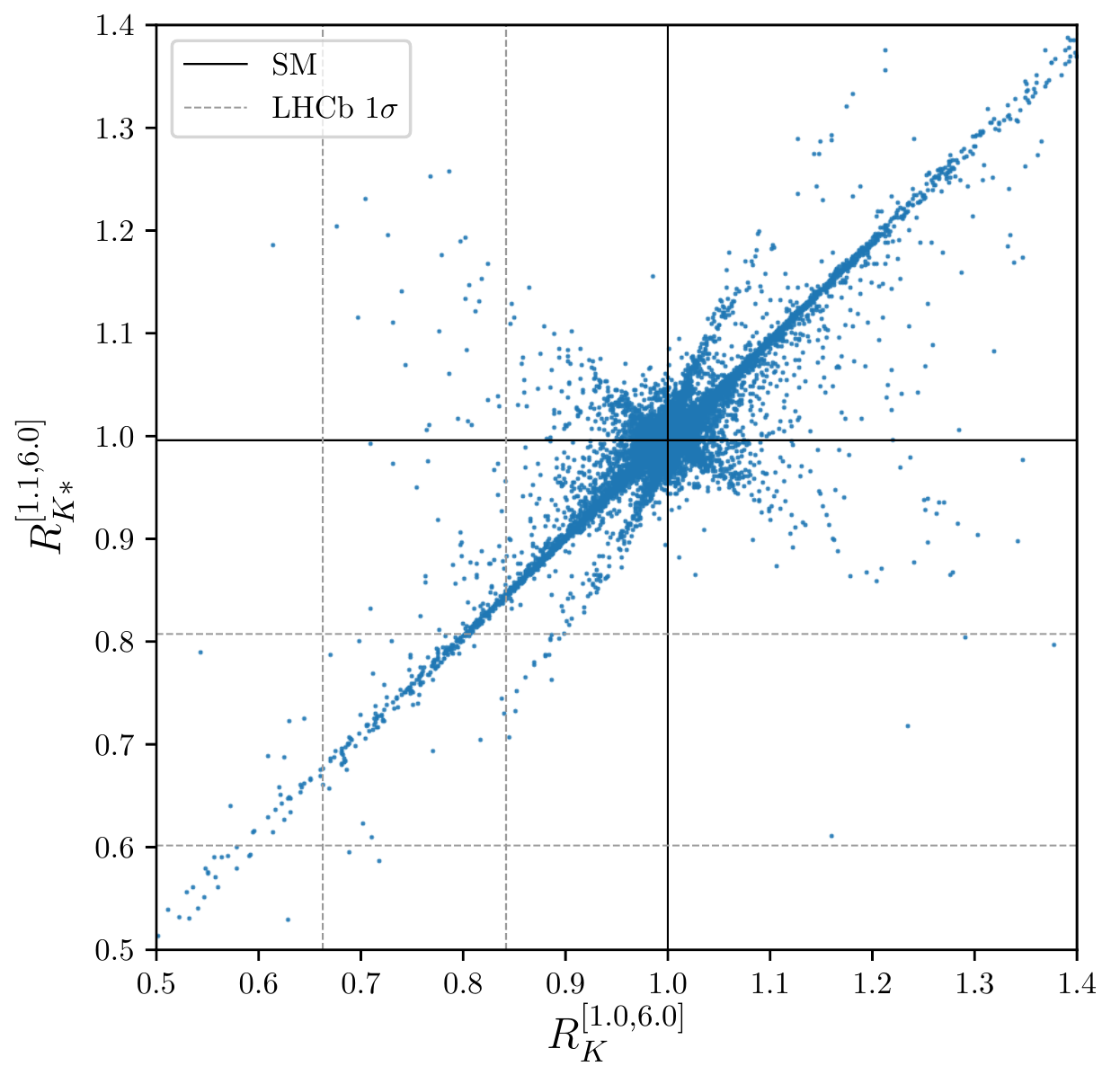}
\quad	\ \	\includegraphics[width=0.34\textwidth]{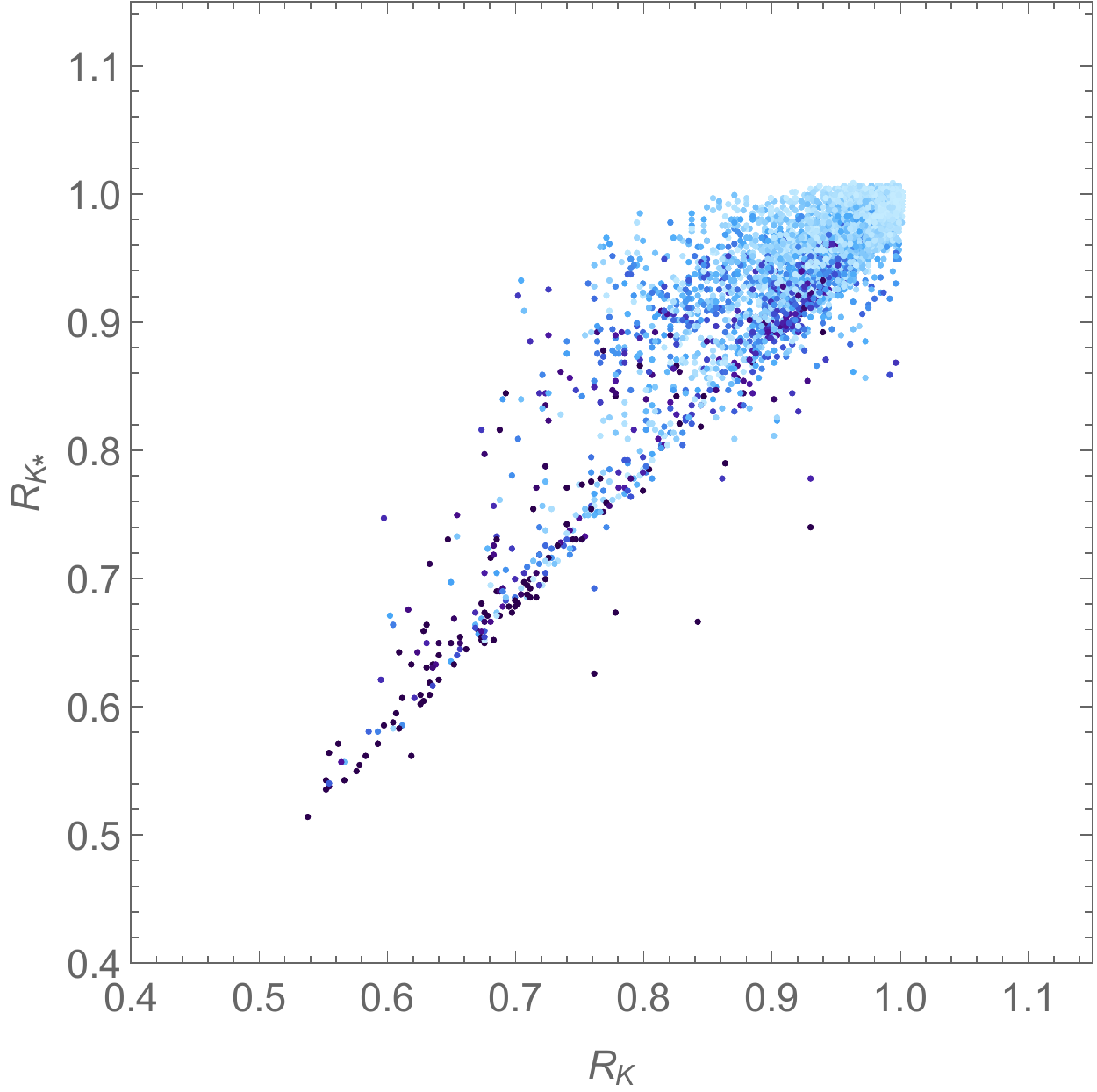} \hspace{-2.2mm}\raisebox{5mm}{
\includegraphics[width=0.0845\textwidth]{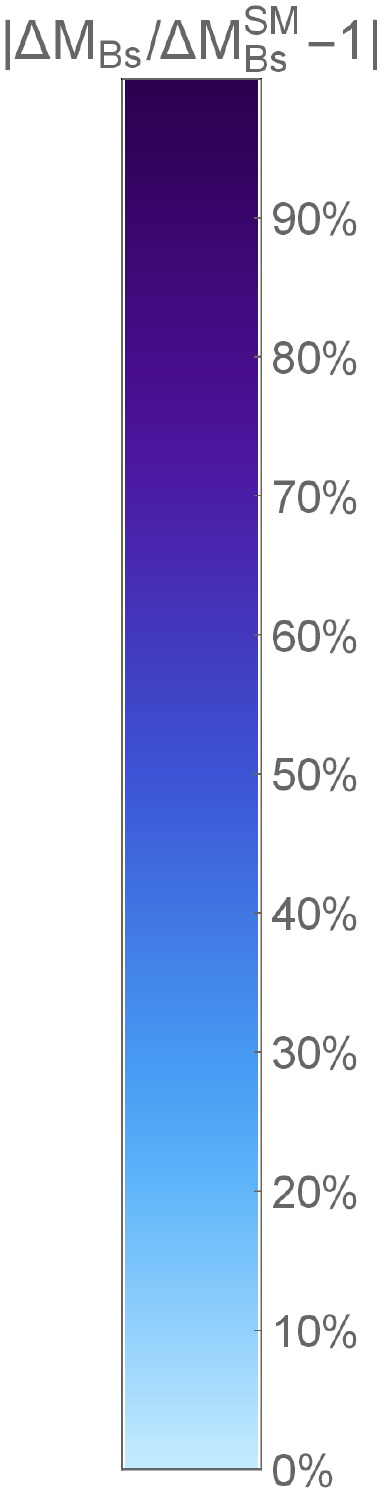}}
		  \vspace*{-3.2mm} \caption{\footnotesize Left: MFPC prediction in the $R_K\!-\!R_{K^\ast}$ plane for 1\,TeV$<\!f\!<3$\,TeV (from \cite{Sannino:2017utc}).  Right: Corresponding prediction (including deviations in the $B_s$ mass difference) in the minimal seesaw model~\cite{Carmona:2017fsn} with $f=1.2\,$TeV.}
		   \label{fig:RK}
		\end{center}
		\vspace{-5.9mm}
\end{figure}

\vspace{-1mm}
\paragraph{Lepton Flavor Universality}
Moving to LFU-violating observables, we focus on the prominent ratios $R_{K^{\!(\ast)}} \!\equiv\! {\rm BR}(B\!\to\! K^{(\ast)} \mu^+\mu^-)/{\rm BR}(B\! \to\! K^{(\ast)} e^+e^-)$, which have been constrained at LHCb as $R_K\!=\!0.846^{+0.060}_{-0.054}$ and $R_{K^\ast}\!=\!0.69^{+0.11}_{-0.07}$ \cite{Aaij:2019wad,Aaij:2017vbb} (for $1.1\!\!<\!q^2\!/{\rm GeV}^2\!\!<\!6$), strongly disfavoring $R_{K^{\!(\ast)}}\!>\!1$, where in the SM $R_K\!\approx\! R_{K^\ast}\!\approx\! 1$. In CH models, corrections are typically induced via four-fermion operators due to heavy EW resonance exchange~\cite{
Niehoff:2015bfa,Carmona:2015ena,Megias:2016bde,Megias:2017ove,Carmona:2017fsn,Sannino:2017utc,Frigerio:2018uwx}, reading (with $X,\!Y\!=\!L,\!R$)
\vspace*{-0.35mm}
\begin{equation}
\mathcal{O}_{XY}^{q^1\!q^2 \ell^1\!\ell^1} = (\bar{q}^1_X\gamma_{\mu} q^2_X)(\bar{\ell}^1_Y\gamma^{\mu} \ell^2_Y)\,,\quad  \text{with coefficients \ } C_{XY}^{q^1\!q^2 \ell^1\!\ell^2}\!\!\sim g_\ast^2/m_\ast^2\ \epsilon_X^{q^1}\epsilon_X^{q^2}\epsilon_Y^{\ell^1}\epsilon_Y^{\ell^2}\,.
\vspace{-0.6mm}
\end{equation}
The LHCb results already notably constrain $\mu_R$ compositeness (for some $\epsilon_X^{s}\epsilon_X^{b}\!>\!0$) and $\mu_L/e_L$~compositeness together with a non-negligible $\epsilon_R^{s}\epsilon_R^{b}\!>\!0$~\cite{Carmona:2017fsn}:
reaching $R_{K,K^\ast}\!<\!1$ requires $\mu_L$~or~$e_L$~compositeness and basically LH $b\!-\!s$ compositeness {\it or} $e_R$  compositeness, irrespectively of the $b\!-\!s$ chirality~\cite{Carmona:2017fsn} (see also \cite{Niehoff:2015bfa,DAmico:2017mtc}). Focusing on LH muons\! / \!RH electrons, we find a good~fit~\cite{Altmannshofer:2017yso,Aebischer:2019mlg,DAmico:2017mtc}~for
\vspace*{-3.4mm}
\begin{equation}
g_\ast^2/m_\ast^2\ \epsilon_L^{s}\epsilon_L^{b}\epsilon_L^{\mu}\epsilon_L^{\mu} \sim 10^{-3}/{\rm TeV}^2 \quad {\rm or} \quad 
g_\ast^2/m_\ast^2\ \epsilon_X^{s}\epsilon_X^{b}\epsilon_R^{e}\epsilon_R^{e} \sim 4 \cdot\!  10^{-3}/{\rm TeV}^2\,.
\vspace*{-0.15mm}
\end{equation} 

Trying to realize one of these patterns led to some efforts in the CH community, envisaging in particular enhanced muon compositeness \cite{Megias:2016bde,Megias:2017ove,Niehoff:2015bfa,Sannino:2017utc,Frigerio:2018uwx}.\footnote{Note that the $R_{D^{(\ast)}}$ anomalies could also be tackled in CH via leptoquark states~\cite{Barbieri:2016las,Blanke:2018sro,Fuentes-Martin:2020bnh}, see also~\cite{Angelescu:2021nbp,Angelescu:2021qbr}.}
In {\it MFPC}, this corresponds to sizable $(y_L)_{22}$ in (\ref{eq:PCF}) (assuming approximate diagonality), while in turn $(y_{\bar e})_{22}$ is tiny to fit the muon mass. After imposing important constraints on this scenarios from 
${\rm BR}(K^+ \to e^+ \nu)/{\rm BR}(K^+ \to \mu^+ \nu)$ and the $Z$-boson partial width (being particular relevant for MFPC which lacks custodial protection), the predictions in the $R_K - R_{K^\ast}$ plane are shown in the left panel of Fig.~\ref{fig:RK}~\cite{Sannino:2017utc}. The best fit value is in principle reachable, and similar conclusions were obtained in ({\it not-quite anarchic}) holographic \cite{Megias:2016bde,Megias:2017ove} and 4D effective~\cite{Niehoff:2015bfa} setups. It would then be interesting to find a motivated model that makes a clearer prediction for LFU violation.
This is provided by the {\it minimal seesaw model}, which predicts moderate RH electron compositeness leading strictly to $R_{K,K^\ast}<1$ -- see the right panel of Fig.~\ref{fig:RK}~\cite{Carmona:2017fsn}. A good fit of $R_{K,K^\ast}\!\sim\! 0.8$ is obtained relatively easily, while respecting other constraints due to small LH lepton compositeness and custodial protection of RH $Z$ couplings.

In conclusion, the different patterns of predictions discussed above, combined with searches for (light) resonances, offer a promising means to get a better handle on the nature~of~leptons.


\section*{Acknowledgements}
\vspace{-4mm}
\noindent
I would like to thank the organizers for the opportunity to present this work at PANIC2021.
\vspace{-0.5mm}

\bibliographystyle{hunsrt}
\footnotesize
\bibliography{CL}

\end{document}